\documentclass[prd,twocolumn]{revtex4}
\usepackage{graphicx, epsfig}
\usepackage{color}
\usepackage{mathrsfs}
\usepackage{amsmath}

\newcommand{\be}{\begin{equation}}
\newcommand{\ee}{\end{equation}}
\newcommand{\bd}{\begin{equation*}}
\newcommand{\ed}{\end{equation*}}
\newcommand{\bea}{\begin{eqnarray}}
\newcommand{\eea}{\end{eqnarray}}

\newcommand{\gapp}{\mathrel{\raise.3ex\hbox{$>$}\mkern-14mu
              \lower0.6ex\hbox{$\sim$}}}
\newcommand{\lapp}{\mathrel{\raise.3ex\hbox{$<$}\mkern-14mu
              \lower0.6ex\hbox{$\sim$}}}

\begin{document}

\title{Time Evolution of Entropy of a Charged Domain Wall during Gravitational Collapse}
\author{Eric Greenwood}
\affiliation{CERCA, Department of Physics, Case Western Reserve University, Cleveland, OH 44106-7079}
\begin{abstract}
We study the time evolution of the entropy of a collapsing charged spherical domain wall, from the point of view of an asymptotic observer, by investigating the entropy of the entire system (i.e.~charged domain wall and radiation) and induced radiation alone during the collapse. By taking the difference, we find the entropy of the collapsing charged domain wall, since this is the object which will form a black hole. We find that for large values of time (times larger than $tf_-/R_+\approx8$), the entropy of the collapsing domain wall is a constant, which is of the same order as the Bekenstein-Hawking entropy. 
\end{abstract}


\maketitle
\section{Introduction}

Hawking's 1974 paper (see Refs.~\cite{Hawking,Hartle}) sparked great interest in both the existence of the radiation given off during the loss of mass for a pre-existing static black hole and speculations as to what his result implied for the validity of quantum field theory in the presence of a black hole. Hawking showed that the radiation given off was in fact thermal, which only increased the interest.

In 1972 Bekenstein argued that a black hole of mass $M$ has an entropy proportional to its area (see Ref.~\cite{Bekenstein}). Further calculations by Gibbons and Hawking showed that the entropy of a black hole is always a constant, regardless of the type of metric which is used (see Ref.~\cite{GibbonsHawking}). They showed that the expression for the entropy is given by 
\be
  S_{BH}=\frac{A_{hor}}{4}
  \label{SBH}
\ee
where $A_{hor}$ is the surface area of the horizon.

Recently, subsequent work has been done in theories of quantum gravity (such as String Theory and Quantum Loop Gravity) to reproduce this result (see for example Refs.~\cite{loopQG,ST}). So far, most of these calculations have involved only pre-existing black holes. The typical method for finding the entropy of the black hole is to first determine the temperature of the black hole using the Bogolyubov method. Here, one considers that the system starts in an asymptotically flat metric, then the system evolves to a new asymptotically flat metric. By matching the coefficients between the two asymptotically flat spaces at the beginning and end of the gravitational collapse, the mismatch of these two vacua gives the number of particles produced during the collapse. Hence, what happens in between is beyond the scope of the Bogolyubov method. Therefore the time-evolution of the thermodynamic properties of the collapse cannot be investigated in the context of the Bogolyubov method. 

In this paper we will investigate the collapse of a charged spherically symmetric infinitely thin domain wall (representing a shell of charged matter). The purpose of this paper is investigate the time-dependent entropy evolution of the charged domain wall during gravitational collapse and verify the claim that the entropy of the black hole is independent of the background metric. Therefore we will compare the results to Eq.~(\ref{SBH}). The details about the collapse will depend on the particular foliation of space-time used to study the system. For the purposes of the question we are studying, we are outside asymptotic observers. Thus, we study the collapse of the domain wall from the view point of a stationary asymptotic observer.  

In order to study the time-dependent thermodynamic processes of the system, we will repeat the steps first developed in Ref.~\cite{Green_ent} for the massive case. This is done by first developing the partition function, in Section \ref{part_func}. To do so we shall employ the so-called Liouville-von Neumann approach (see Refs.~\cite{Thermal,Kim}). This approach utilizes the invariant operator approach developed by Lewis and Reisenfeld (see Ref.~\cite{Lewis}). In the invariant operator formalism, it was observed that any Hermitian operator $I(t)$ satisfying the Liouville-von Neumann equation,
\be
  \frac{dI}{dt}=\frac{\partial I}{\partial t}-i\left[I,H\right]=0
  \label{LvN}
\ee
can be used to find the exact wavefunctions of the Schr\"odinger equation, up to an overall time-dependent phase factor. In the Liouville-von Neumann approach, the time-dependent Hamiltonian is replaced by the time-dependent invariant operator, since the eigenvalues of this operator are time-independent and all the time-dependence of the system is placed into a non-linear auxiliary equation. 

To find the parameter $\beta$, we shall use the so-called Functional Schr\"odinger formalism (see for example Refs.~\cite{Stojkovic,Greenwood,WangGreenStoj}), in Section \ref{sec:beta}. In general, the Functional Schr\"odinger formalism yields a functional differential equation for the wavefunctional, $\Psi[g_{\mu\nu},X^{\mu},\Phi,{\cal{O}}]$, where $g_{\mu\nu}$ is the metric, $X^{\mu}$ the position of the object, $\Phi$ is a scalar field and ${\cal{O}}$ denotes all the observer's degrees of freedom. Since the Functional Schr\"odinger formalism depends on the observer's degrees of freedom, one can introduce the ``observer" time into the quantum mechanical processes in the form of the Schr\"odinger equation. The wavefunctional is then dependent on the chosen observer's time, hence one can view the quantum mechanical processes of a given system under any choice of space-time foliation. One benefit of this formalism is that one can solve the time-dependent wavefunctional equation exactly using the invariant operator method, in particular  in the present paper. Here we study the propagation of induced radiation, represented as a scalar field in the background of gravitational collapse, which is governed by the harmonic oscillator equation with a time-dependent frequency. We solve the equations of motion to find the time-dependent wavefunctional for the system. One can then define the occupation number of the induced radiation using the wavefunctional, which is the gaussian overlap between the initial vacuum state and the wavefunction at any given later time. By taking the occupation number to be of the form of the Planck distribution, we can then define $\beta$ which is also time-dependent. 

Since the partition function is time-dependent, one can then study the time-dependent thermodynamic processes associated with gravitational collapse. Therefore the Functional Schr\"odinger formalism and Liouville-von Neumann approach together go beyond the approximations of the Bogolyubov method, since the system is allowed to evolve over time, which allows one to investigate the intermediate regime during the collapse.
 
\section{Model for Collapsing Charged Domain Wall}
\label{Model}

To study a realization of the formation of a black hole, we consider a spherical Nambu-Goto domain wall (representing a domain wall of matter) that is collapsing. This will be done using the Functional Schr\"odinger equation, where we will consider a minisuperspace version of the Wheeler-de Witt equation. Since we are interested in the entropy of the collapsing domain wall from the point of view of an asymptotic observer, the relevant degree of freedom for the collapsing spherical domain wall is the radial degree of freedom $R(t)$. The metric of the system is then chosen to be the solution to Einstein's equations for a spherical domain wall. Therefore, outside the wall the metric is Schwarzschild, as follows from the spherical symmetry
\begin{align}
  ds^2=&-\left(1-\frac{2GM}{r}+\frac{Q^2}{r^2}\right)dt^2\nonumber\\
        &+\left(1-\frac{2GM}{r}+\frac{Q^2}{r^2}\right)^{-1}dr^2+r^2d\Omega^2
  \label{metric}
\end{align}
where  
\be
  d\Omega^2=d\theta^2+\sin^2\theta d\phi^2.
\ee
By Birkhoff's theorem, the metric inside the domain wall must be flat, hence Minkowski
\be
  ds^2=-dT^2+dr^2+r^2d\Omega^2.
  \label{metricout}
\ee
where $T$ is the interior time. The interior time is related to the asymptotic observer time $t$ via the proper time $\tau$ of the domain wall.
\be
  \frac{dT}{d\tau}=\sqrt{1+\left(\frac{dR}{d\tau}\right)^2}
  \label{dTdtau}
\ee 
and
\be
  \frac{dt}{d\tau}=\frac{1}{f}\sqrt{f+\left(\frac{dR}{d\tau}\right)^2}
  \label{dtdtau}
\ee
where
\be
  f\equiv1-\frac{2GM}{R}+\frac{Q^2}{R^2}.
  \label{f}
\ee
By taking the ratio of these equations we then have
\be
  \frac{dT}{dt}=\sqrt{f-\frac{(1-f)}{f}\dot{R}^2}
  \label{dTdt}
\ee
where $\dot{R}=dR/dt$.

By integrating the equation of motion for a spherically symmetric domain wall, Wang \textit{et.~al.} and L\'opez (see Refs.~\cite{WangGreenStoj,Lopez}) found that the mass is a constant of motion and is given by
\be
  M=4\pi\sigma R^2\left[\sqrt{1+R_{\tau}^2}-2\pi\sigma GR\right]+\frac{Q^2}{2R}
  \label{IpMass}
\ee
where $\sigma$ is the surface tension of the domain wall and $R_{\tau}=dR/d\tau$. Before we proceed we wish to discuss the physical relevance of Eq.~(\ref{IpMass}). First consider the case where $R_{\tau}=0$, i.e. for a static domain wall. The first term in the square bracket is just the total rest mass of the domain wall. When the domain wall is moving, i.e. $R_{\tau}\not=0$, the first term in the square bracket takes the kinetic energy of the domain wall into account. The last term in the square bracket is the self-gravity, or the binding energy of the domain wall. Finally, the last term in Eq.~(\ref{IpMass}) is the electromagnetic contribution to the total mass (energy). Therefore we can identify Eq.~(\ref{IpMass}) as the total energy of the system, hence the Hamiltonian of the system. Thus, we will refer to Eq.~(\ref{IpMass}) as the Hamiltonian.

Using Eq.~(\ref{dtdtau}) the Hamiltonian for the wall, Eq.~(\ref{IpMass}), can be written as (see Ref.~\cite{WangGreenStoj})
\be
  H=4\pi\sigma f^{3/2}R^2\left[\frac{\frac{Q^2}{8\pi\sigma R^3}-2\pi\sigma GR}{\sqrt{f^2-(1-f)\dot{R}^2}}-\frac{1}{\sqrt{f^2-\dot{R}^2}}\right]
  \label{mass}
\ee
where $\dot{R}=dR/dt$. The canonical momentum near the horizon, the $R\sim R_H$ regime, is given by
\be
  \Pi_R\approx\frac{4\pi\mu R^2\dot{R}}{\sqrt{f}\sqrt{f^2-\dot{R}^2}}
\ee
where 
\be
  \mu\equiv1-2\pi\sigma GR_H+\frac{Q^2}{8\pi\sigma R_H^3}. 
\ee
In this region the wall Hamiltonian Eq.~(\ref{mass}) in terms of the canonical momentum is then
\bea
  H_{wall}&\approx&\frac{4\pi\mu f^{3/2}R^2}{\sqrt{f^2-\dot{R}^2}}\nonumber\\
     &=&\sqrt{(f\Pi_R)^2+f(4\pi\mu R^2)^2}
     \label{massHam}
\eea
The Hamiltonian has the form of the energy of a relativistic particle, $\sqrt{p^2+m^2}$, with a position-dependent mass.

In the region $R\sim R_H$, Eq.~(\ref{f}) tells us that $f\sim0$. From Eq.~(\ref{massHam}) one can see that the mass term can be neglected, thus Eq.~(\ref{massHam}) reduces to
\be
  H_{wall}\approx-f\Pi_R
  \label{HW}
\ee
where we have chosen the negative sign since the domain wall is collapsing. 

Considering only the classical solution, the near-horizon solution of Eq.~(\ref{mass}) can be written as
\be
  \dot{R}\approx-f=-\left(1-\frac{2GM}{R}+\frac{Q^2}{R^2}\right).
  \label{Rdot}
\ee
For the nonextremal case, we consider the equality
\be
  1-\frac{2GM}{R_H}+\frac{Q^2}{R^2_H}=0,
\ee
where $R_H$ is given by
\be
  R_H=GM\pm\sqrt{(GM)^2-Q^2}.
\ee
The plus sign is the outer and the minus sign is the inner horizon. To distinguish them we write
\be
  R_+=GM+\sqrt{(GM)^2-Q^2}
  \label{Rplus}
\ee
\be
  R_-=GM-\sqrt{(GM)^2-Q^2}
\ee
Therefore we can write Eq.~(\ref{f}) as
\begin{align}
  f&=\left(1-\frac{R_+}{R}\right)\left(1-\frac{R_-}{R}\right)\nonumber\\
   &\equiv f_+f_-.
\end{align}
Since in the near horizon limit $f\rightarrow0$, we can work with two different limits, either $f_+\rightarrow0$ or $f_-\rightarrow0$. In both cases $\dot{R}\approx\pm f_+f_-\rightarrow0$. For an asymptotic observer watching the collapse, the horizon of interest is $f_+$. For $f_+\rightarrow0$, $f_-$ goes to a finite constant number, thus we have that the solution for Eq.~(\ref{Rdot}) is then
\be
  R(t)\approx R_++(R_0-R_+)e^{-f_-t/R_+}.
  \label{classical}
\ee
Eq.~(\ref{classical}) implies that a collapsing domain wall crosses its own horizon only after an infinite amount of asymptotic observer time $t$. 

Here we note that Eq.~(\ref{classical}) implies that in the case $Q^2>(GM)^2$, the argument in the exponential becomes imaginary. Thus the solution for the collapsing charged domain wall contains sinusoidal terms which oscillate about a radius $\tilde{R}$ outside of the horizon. In this case the horizon never forms, thus never creating a black hole. Therefore, we shall ignore this solution and only investigate solutions where $(GM)^2>Q^2$. 

\section{Induced Radiation}
\label{Radiation}

Here we consider the radiation given off during gravitational collapse. We will consider the radiation given off by the entire system and the particles produced during collapse alone. 

To investigate the radiation, we consider a complex scalar field $\Phi$ in the background of the collapsing domain wall. The complex scalar field is decomposed into a complete set of basis functions denoted by $\{f_k(r)\}$
\be
  \Phi=\sum_k\left(a_k(t)f_k(r)+ib_k(t)f_k(r)\right).
  \label{expansion}
\ee
The exact form of the functions $f_k(r)$ will not be important for us. We will, however, be interested in the wavefunction for the mode coefficients $\{a_k\}$ and $\{b_k\}$.

The Hamiltonian for the complex scalar field modes is found by inserting Eq.~(\ref{expansion}) and Eq.~(\ref{metric}) and Eq.~(\ref{metricout}) into the action
\be
  S_{\Phi}=\int d^4x\sqrt{-g}\frac{1}{2}g^{\mu\nu}(D_{\mu}\Phi)^*(D_{\nu}\Phi)
  \label{rad act}
\ee
where
\be
  D_{\mu}=\partial_{\mu}+iqA_{\mu}
\ee
is the covariant derivate, $q$ is the charge of the field and $A_{\mu}$ is the vector potential. However, from Ref.~\cite{Israel}, we can write the vector potential for the collapsing charged shell as 
\be
  (A_{\mu})=\left(\frac{Q}{r},0,0,0\right).
\ee
The Hamiltonian for the complex scalar field modes, arrived at from Eq.~(\ref{rad act}), takes the form of uncoupled simple harmonic oscillators for each mode with $R$-dependent mass and couplings due to the non-trivial metric. Using the principle axis transformation, the Hamiltonian can be diagonalized and written in terms of eigenmodes denoted by $c$ and $d$, where $c$ is the eigenmode for $a$ and $d$ is the eigenmode for $b$. It was shown in Ref.~\cite{Green_rad} that in the regime of $R\sim R_+$, the Hamiltonian for a single $c$ mode can be written as
\be
  H_{c}=f\frac{\Pi_c^2}{2m}+i\frac{qQy}{2m}+\frac{K}{2}c^2
  \label{Hrad_c}
\ee
and for a single $d$ mode
\be
  H_{d}=f\frac{\Pi_d^2}{2m}-i\frac{qQy}{2m}+\frac{K}{2}d^2
  \label{Hrad_d}
\ee
where $\Pi_c$ is the momentum conjugate to $c$, $\Pi_d$ is the conjugate momentum to $d$, and where $m$, $y$ and $K$ are constants whose precise values are not important to us.  

The physical meaning of the $c$ and $d$ modes are inferred from the Noether current. From Ref.~\cite{Green_rad}, we can write the Noether current as
\be
  N_{\Phi}=\frac{m}{f}\left(\dot{c}c-\dot{d}d\right)+qQy\left(c^2+d^2\right)
  \label{NC}
\ee
where $\dot{c}=dc/dt$ and $\dot{d}=dd/dt$. The mode coefficients satisfy the commutation relations given by
\be
  [\dot{c},c]=[\dot{d},d]=1,
\ee
hence we see that the the Noether current satisfies the commutation relations
\be
  \left[N_{\Phi},c\right]=c, \hspace{1mm} \text{and} \hspace{1mm} \left[N_{\Phi},d\right]=-d.
\ee
Therefore we see that the $c$ modes create particles with $+q$ worth of charge, while the $d$ modes create particles with $-q$ worth of charge. 

\subsection{Entire System}

From Eq.~(\ref{HW}) and Eqs.~(\ref{Hrad_c}) and (\ref{Hrad_d}) we can write the Hamiltonian of the entire system as
\begin{align}
  H&=H_{wall}+H_b\nonumber\\
      &=-f\Pi_R+f\frac{\Pi_c^2}{2m}+i\frac{qQy}{2m}+\frac{K}{2}c^2+f\frac{\Pi_d^2}{2m}-i\frac{qQy}{2m}+\frac{K}{2}d^2\nonumber\\
      &=-f\Pi_R+f\frac{\Pi_c^2}{2m}+\frac{K}{2}c^2+\frac{K}{2}d^2
  \label{tot Ham}
\end{align}
where $\Pi_R=-i\partial/\partial R$, $\Pi_c=-i\partial/\partial c$ and $\Pi_d=-i\partial/\partial d$. As pointed out in Ref.~\cite{Green_rad}, the total wavefunction is independent of the charge charge of the modes. The wavefunction for the entire system is then a function of $c$, $d$, $R$, and $t$, which we can write as
\be
  \Psi=\Psi(c,d,R,t).
\ee

Using the standard quantization procedure and Eq.~(\ref{tot Ham}), we can then write the Functional Schr\"odinger equation as
\be
  if\frac{\partial\Psi}{\partial R}-\frac{f}{2m}\frac{\partial^2\Psi}{\partial c^2}-\frac{f}{2m}\frac{\partial^2\Psi}{\partial d^2}+\frac{K}{2}c^2\Psi+\frac{K}{2}d^2\Psi=i\frac{\partial\Psi}{\partial t}.
  \label{Schrod1}
\ee
To solve Eq.~(\ref{Schrod1}) we choose the classical background for the collapsing domain wall. Since the distance of the domain wall only depends on time, see Eq.~(\ref{Rdot}), we can then write
\begin{equation*}
  if\frac{dt}{dR}\frac{\partial\Psi}{\partial t}-\frac{f}{2m}\frac{\partial^2\Psi}{\partial c^2}-\frac{f}{2m}\frac{\partial^2\Psi}{\partial d^2}+\frac{K}{2}c^2\Psi+\frac{K}{2}d^2\Psi=i\frac{\partial\Psi}{\partial t}.
  \label{Schrod2}
\end{equation*}
Subtracting the first time from both sides and grouping like terms, we can then write Eq.~(\ref{Schrod1}) as
\be
  -\frac{f}{2m}\frac{\partial^2\Psi}{\partial c^2}-\frac{f}{2m}\frac{\partial^2\Psi}{\partial d^2}+\frac{K}{2}c^2\Psi+\frac{K}{2}d^2\Psi=i\frac{\partial\Psi}{\partial t}\left(1-f\frac{dt}{dR}\right).
  \label{Schrod3}
\ee
Making use of the classical equation of motion, Eq.~(\ref{Rdot}), Eq.~(\ref{Schrod3}) becomes
\be
  -\frac{f}{2m}\frac{\partial^2\Psi}{\partial c^2}-\frac{f}{2m}\frac{\partial^2\Psi}{\partial d^2}+\frac{K}{2}c^2\Psi+\frac{K}{2}d^2\Psi=2i\frac{\partial\Psi}{\partial t}.
  \label{ent Schrod}
\ee
Thus we treat the background classically and the radiation quantum mechanically.

Dividing Eq.~(\ref{ent Schrod}) through by $f$, we now rewrite in the standard form of a harmonic oscillator
\begin{align}
  &\left[-\frac{1}{2m}\frac{\partial^2}{\partial c^2}-\frac{1}{2m}\frac{\partial^2}{\partial d^2}+\frac{m}{2}\omega^2(\eta)c^2+\frac{m}{2}\omega^2(\eta)d^2\right]\psi(c,d,\eta)\nonumber\\
    &=i\frac{\partial\psi(c,d,\eta)}{\partial \eta}
  \label{Ent Schrod}
\end{align}
where
\be
  \eta=\frac{1}{2}\int_0^tdt' f=\frac{1}{2}\int_0^tdt'\left(1-\frac{2GM}{R}+\frac{Q^2}{R^2}\right)
\ee
and
\be
  \omega^2(\eta)=\frac{K}{m}\frac{1}{1-R_s/R+Q^2/R^2}\equiv\frac{\omega_0^2}{1-R_s/R+Q^2/R^2}
\ee
where $\omega_0^2\equiv K/m$. Here we have chosen to set $\eta(t=0)=0$.

To solve Eq.~(\ref{Ent Schrod}), we make the ansatz that the wave-function is separable, hence we can write the total wave-function as
\be
  \Psi(c,d,\eta)=\psi(c,\eta)\psi(d,\eta).
  \label{wave_sep}
\ee
Therefore each of the mode equations satisfies the time-dependent harmonic oscillator equation separately. The solution for each mode in Eq.~(\ref{Ent Schrod}) is given by (see Ref.~\cite{Pedrosa})
\be
  \psi_{S,R}(c,\eta)=e^{i\alpha(\eta)}\left(\frac{m}{\pi\rho^2}\right)^{1/4}\exp\left[\frac{im}{2}\left(\frac{\rho_{\eta}}{\rho}+\frac{i}{\rho^2}\right)c^2\right]
  \label{wave func_c}
\ee
and
\be
  \psi_{S,R}(d,\eta)=e^{i\alpha(\eta)}\left(\frac{m}{\pi\rho^2}\right)^{1/4}\exp\left[\frac{im}{2}\left(\frac{\rho_{\eta}}{\rho}+\frac{i}{\rho^2}\right)d^2\right]
  \label{wave func_d}
\ee
where $\rho_{\eta}=d\rho/d\eta$ is the derivative of the function $\rho(\eta)$ with respect to $\eta$, and $\rho$ is the real solution to the non-linear auxiliary equation
\be
  \rho_{\eta\eta}+\omega^2(\eta)\rho=\frac{1}{\rho^3}
  \label{rho}
\ee
and where $\alpha$ is the phase, which is given by
\be
  \alpha(\eta)=-\frac{1}{2}\int^{\eta}\frac{d\eta'}{\rho^2(\eta')}.
\ee

\subsection{Radiation Only}

In this section we only consider the particles created during the collapse of the domain wall. To describe this we use only Eqs.~(\ref{Hrad_c}) and (\ref{Hrad_d}). In this case the Functional Schr\"odinger equation becomes
\be
  -\frac{f}{2m}\frac{\partial^2\Psi}{\partial c^2}-\frac{f}{2m}\frac{\partial^2\Psi}{\partial d^2}+\frac{K}{2}c^2\Psi+\frac{K}{2}d^2\Psi=i\frac{\partial\Psi}{\partial t}.
  \label{rad Schrod}
\ee
Dividing through by $f$ and writing in the standard form yields
\be
  -\frac{1}{2m}\frac{\partial^2\Psi}{\partial c^2}-\frac{1}{2m}\frac{\partial^2\Psi}{\partial d^2}+\frac{m}{2}\omega^2(\tilde{\eta})c^2\Psi+\frac{m}{2}\omega^2(\tilde{\eta})d^2\Psi=i\frac{\partial\Psi}{\partial \tilde{\eta}}
  \label{Rad Schrod}
\ee
where
\be
  \tilde{\eta}=\int_0^tdt' f=\int_0^tdt'\left(1-\frac{R_s}{R}\right)
  \label{tilde_eta}
\ee
and 
\be
  \omega^2(\tilde{\eta})=\frac{\omega^2_0}{1-R_s/R+Q^2/R^2}.
\ee
where $\omega_0^2$ is defined as before. Again separating the wave-function as in Eq.~(\ref{wave_sep}), the solution to Eq.~(\ref{Rad Schrod}) is the same as Eqs.~(\ref{wave func_c}) and (\ref{wave func_d}) except with the replacement $\eta$ now given by $\tilde{\eta}$ (Eq.~(\ref{tilde_eta})), i.e.
\be
  \psi_{R}(c,\tilde{\eta})=e^{i\alpha(\tilde{\eta})}\left(\frac{m}{\pi\rho^2}\right)^{1/4}\exp\left[\frac{im}{2}\left(\frac{\rho_{\tilde{\eta}}}{\rho}+\frac{i}{\rho^2}\right)c^2\right]
  \label{Wave func_c}
\ee
and
\be
  \psi_{R}(d,\tilde{\eta})=e^{i\alpha(\tilde{\eta})}\left(\frac{m}{\pi\rho^2}\right)^{1/4}\exp\left[\frac{im}{2}\left(\frac{\rho_{\tilde{\eta}}}{\rho}+\frac{i}{\rho^2}\right)d^2\right]
  \label{Wave func_d}
\ee
Hence each wavefunction evolves with a different time parameter.

\subsection{Taking the Temperature}
\label{sec:beta}

For an observer, the complete information about the radiation, in the background of the collapsing domain wall, is contained in the wavefunction. Consider an observer with detectors that are designed to register particles of different frequencies for the free field $\Phi$. Such an observer will interpret the wave function of a given mode at some late time in terms of simple harmonic oscillator states.

For brevity, we consider the following analysis using the notation for the entire system. However, the analysis for the radiation alone is the same.

The wave functions in Eqs.~(\ref{wave func_c}) and (\ref{wave func_d}) (Eqs.~(\ref{Wave func_c}) and (\ref{Wave func_d})) can be decomposed into suitably chosen vacuum basis wave functions, denoted by $\{\phi_n\}$ at the final frequency $\bar{\omega}$. The number of quanta in eigenmodes $c$ and $d$ can be evaluated by decomposing the wavefunction in terms of the states $\{\varphi_n\}$ and by evaluating the occupation number of that mode. We decompose the wavefunction at a time $t>t_f$ in terms of a simple harmonic oscillator basis at $t=0$. The decomposition is 
\be
  \Psi(c,d,t)=\sum_{n,m}c_n(t)c_m(t)\varphi_n(c)\varphi_m(d)
\ee
where
\be
  c_n=\int dc\varphi_n^*(c)\psi(c,t) \hspace{2mm} \text{and} \hspace{2mm} c_m=\int dd\varphi_m^*(d)\psi(d,t)
  \label{c_n}
\ee
which are the overlaps of a Gaussian with the $t=0$ simple harmonic oscillator basis functions. It was shown in Ref.~\cite{Green_rad} that the expectation values of the occupation number at given eigenfrequency $\bar{\omega}$ (i.e.~in the eigenmodes $c$ and $d$) by the time $t>t_f$, are then given by
\be
  N_c(t,\bar{\omega})=\sum_nn\left|c_n\right|^2 \hspace{2mm} \text{and} \hspace{2mm} N_d(t,\bar{\omega})=\sum_mm\left|c_m\right|^2.
  \label{occ}
\ee
The total occupation number is then the sum of the individual occupation numbers, $N=N_c+N_d$. Since the initial state is chosen to be the vacuum state, the number of quanta in eigenmode measured by the occupation number is a consequence of only the collapse, since the observer will initially measure zero quanta. 

As stated previously, the observer will interpret the wavefunction of a given mode in terms of simple harmonic oscillator states. Thus the basis functions $\phi_n$ are chosen to be simple harmonic oscillator basis states at a frequency $\bar{\omega}$
\be
  \phi_n(c)=\left(\frac{m\bar{\omega}}{\pi}\right)^{1/4}\frac{e^{-m\bar{\omega}c^2/2}}{\sqrt{2^nn!}}{\cal{H}}_n\left(\sqrt{m\bar{\omega}}c\right)
  \label{harmonic_phi}
\ee
where ${\cal{H}}_n$ are the Hermite polynomials. For brevity we have only written the simple harmonic oscillator states for the $c$ modes, a similar expression holds true for the $d$ modes as well. Therefore using Eq.~(\ref{c_n}) and Eq.~(\ref{harmonic_phi}) we can write the inner product as in Ref.~\cite{Stojkovic,Greenwood,Green_rad}, for $n=$ odd $c_n=0$ and for $n=$ even
\be
 c_n=\frac{(-1)^{n/2}e^{-i\alpha}}{(\bar{\omega}\rho^2)^{1/4}}\sqrt{\frac{2}{P}}\left(1-\frac{2}{P}\right)^{n/2}\frac{(n-1)!!}{\sqrt{n!}}
  \label{inner prod}
\ee
where
\be
  P\equiv1-\frac{i}{\bar{\omega}}\left(\frac{\rho_{\eta}}{\rho}+\frac{i}{\rho^2}\right).
  \label{P}
\ee
Similar expressions hold true for the $d$ modes as well.

Substituting Eq.~(\ref{inner prod}) and Eq.~(\ref{P}) and performing the sum over $n=$ even in Eq.~(\ref{occ}) we then have for the $c$ modes
\be
  N_c(t,\bar{\omega})=\frac{\bar{\omega}\rho^2}{\sqrt{2}}\left[\left(1-\frac{1}{\bar{\omega}\rho^2}\right)^2+\left(\frac{\rho_{\eta}}{\bar{\omega}\rho}\right)^2\right]
  \label{Occ_c}
\ee
and for the $d$ modes
\be
  N_d(t,\bar{\omega})=\frac{\bar{\omega}\rho^2}{\sqrt{2}}\left[\left(1-\frac{1}{\bar{\omega}\rho^2}\right)^2+\left(\frac{\rho_{\eta}}{\bar{\omega}\rho}\right)^2\right].
  \label{Occ_d}
\ee
Therefore the total occupation number is given as
\be
  N=N_c+N_d=\frac{2\bar{\omega}\rho^2}{\sqrt{2}}\left[\left(1-\frac{1}{\bar{\omega}\rho^2}\right)^2+\left(\frac{\rho_{\eta}}{\bar{\omega}\rho}\right)^2\right]
  \label{Occ}
\ee
by virtue of Eqs.~(\ref{Occ_c}) and (\ref{Occ_d}). By fitting the occupation number to that of the Planck distribution, 
\be
  N_P=\frac{1}{e^{\beta\bar{\omega}}-1}
  \label{NP}
\ee
the occupations at eigenmodes $c$ and $d$ can then be used to find the temperature of the radiation. From Eqs.~(\ref{wave func_c}) and (\ref{wave func_d}) and Eqs.~(\ref{Wave func_c}) and (\ref{Wave func_d}) we see that the occupation number for each of the systems will be different, since both $\eta$ and $\tilde{\eta}$ are different.

\section{Entropy Function}
\label{Entropy}

Here we develop the entropy function for studying the time evolution of the collapsing domain wall. Again for brevity we use the notation for the entire system, where again the analysis is the same for the radiation only system. 

\subsection{Partition Function}
\label{part_func}

To study the entropy of the system, we will first develop the partition function for the system. Following the procedure used in Ref.~\cite{Kim} for the Liouville-von Neumann approach, we can write the partition function as
\be
  Z=\textrm{Tr}\left[e^{-\beta I}\right]
\ee
where $I$ is any operator which satisfies the Liouville-von Neumann equation Eq.~(\ref{LvN}) and $\beta$ is a free parameter. 
 
 From Ref.~\cite{Pedrosa}, we can write the invariant operator $I$ as
\be
  I=\frac{1}{2}\left[\sqrt{\frac{c}{\rho}}+\left(\pi_c\rho-m\rho_{\eta}c\right)^2+\sqrt{\frac{d}{\rho}}+\left(\pi_d\rho-m\rho_{\eta}d\right)^2\right].
\ee
We can therefore write the partition function as
\begin{align}
  Z=\textrm{Tr}\exp\Big{[}&-\beta\frac{1}{2}\Big{[}\sqrt{\frac{c}{\rho}}+\left(\pi_c\rho-m\rho_{\eta}c\right)^2\nonumber\\
  &+\sqrt{\frac{d}{\rho}}+\left(\pi_d\rho-m\rho_{\eta}d\right)^2\Big{]}\Big{]}.
\end{align}

We note that we can rewrite the invariant as
\begin{align}
  I=&\left(\frac{1}{\sqrt{2}}\right)^2\Big{(}\left[\left(\frac{c}{\rho}\right)^{1/4}-i\left(\pi_c\rho-m\rho_{\eta}c\right)\right]\nonumber\\
    &\times\left[\left(\frac{c}{\rho}\right)^{1/4}+i\left(\pi_c\rho-m\rho_{\eta}c\right)\right]\nonumber\\
    &+\left[\left(\frac{d}{\rho}\right)^{1/4}-i\left(\pi_d\rho-m\rho_{\eta}d\right)\right]\nonumber\\
    &\times\left[\left(\frac{d}{\rho}\right)^{1/4}+i\left(\pi_d\rho-m\rho_{\eta}d\right)\right]\Big{)}\nonumber\\
    \equiv&n(t)+m(t)+1
\end{align}
where
\be
  n(t)=a^{\dagger}(t)a(t), \hspace{2mm} \text{and} \hspace{2mm} m(t)=\hat{a}^{\dagger}(t)\hat{a}(t)
\ee
and
\begin{align}
  a(t)&\equiv\frac{1}{\sqrt{2}}\left[\left(\frac{c}{\rho}\right)^{1/4}+i\left(\pi_c\rho-m\rho_{\eta}c\right)\right]\nonumber\\
  \hat{a}(t)&\equiv\frac{1}{\sqrt{2}}\left[\left(\frac{d}{\rho}\right)^{1/4}+i\left(\pi_d\rho-m\rho_{\eta}d\right)\right]
\end{align}
Here $n(t)$ is the time-dependent number of states for the $c$ modes, while $m(t)$ is the time-dependent number of states for the $d$ modes. At a particular time $t$, one has in Fock space
\be
  n(t)\big{|}n,m,t\rangle=n\big{|}n,m,t\rangle, \hspace{2mm} \text{and} \hspace{2mm} m(t)\big{|}n,m,t\rangle=m\big{|}n,m,t\rangle. 
\ee
Thus, in this space we can then write the partition function as
\bea
  Z&=&\textrm{Tr}\exp\left[-\beta\omega_0\left(n+m+1\right)\right]\nonumber\\
   &=&\frac{1}{4\sinh^2\left(\frac{\beta\omega_0}{2}\right)}.
   \label{part func}
\eea
Here we note that Eq.~(\ref{part func}) takes the form of a two-dimensional harmonic oscillator.

In Ref.~\cite{Stojkovic,Greenwood,Green_rad} one can define the occupation number for a frequency $\bar{\omega}$, Eq.~(\ref{Occ}). Then by fitting the number of particles created as the usual Planck distribution Eq.~(\ref{NP}), one can then in principle fit the temperature of the radiation. Here, we then choose to define $\beta$ as 
\be
  \beta=\frac{\partial\ln\left(1+1/N\right)}{\partial\bar{\omega}}.
  \label{beta}
\ee
This implies that all of the time-dependence of the system is encoded into the temperature of the system.

We can see that Eq.~(\ref{part func}) is just the standard entropy for a time-independent harmonic oscillator; however, the temperature here is time-dependent. Therefore the entropy is also time-dependent.

\subsection{Entropy Analysis}
\label{Entropy}

Using Eq.~(\ref{part func}) and the thermodynamic definition of entropy, we can then write the entropy of the system as
\be
  S=-\ln\left((1-e^{-\beta\omega_0})^2\right)+2\beta\frac{e^{-\beta\omega_0}}{1-e^{-\beta\omega_0}}.
  \label{entropy}
\ee
Therefore, this is again just the entropy of the usual two-dimensional time-independent harmonic oscillator. From Eq.~(\ref{beta}) it follows that the temperature is time-dependent. 

\begin{figure}[htbp]
\includegraphics{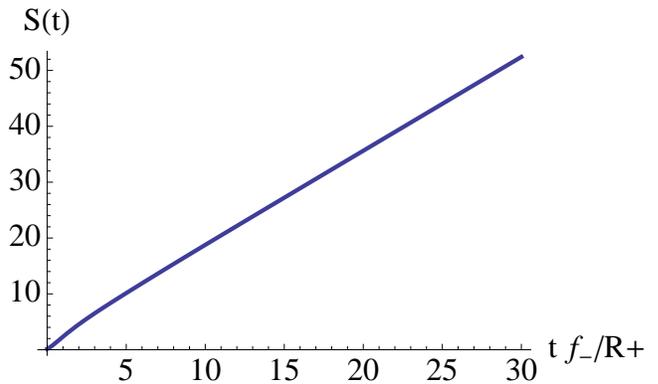}
\caption{We plot the entropy of the entire system as a function of $tf_-/R_+$. Except for the initial increase, the entropy increases approximately linearly as $tf_-/R_+$ increases.}
\label{EntEntire}
\end{figure}

First we consider the entropy of the entire system. In Figure \ref{EntEntire} we plot the entropy of the entire system as a function of dimensionless time $tf_-/R_+$. Figure \ref{EntEntire} shows that the system starts with an initial entropy of zero. This is expected since initially there is only one degree of freedom, meaning that $S=\ln(1)=0$. Here we have normalized the initial entropy of the shell to be zero. To justify this normalization, consider a solar mass black hole. Under the usual Bekenstein-Hawking entropy, the order of magnitude estimate of the entropy of a solar mass black hole is $S_{BH}\approx10^{75}$. Now consider that the shell is actually made up of protons. The initial entropy of the shell then is approximately $S_{S,0}\approx10^{57}$. Comparing the entropy of the final black hole versus the initial entropy of the shell, the entropy of the final black hole is much much greater than that of the initial entropy of the shell, thus the initial entropy of the shell only contributes a negligible amount of entropy to the entropy of the final black hole. Thus our normalization of the initial entropy of the shell to zero is justified. As $tf_-/R_+$ increases, initially the entropy increases rapidly, then settles down to increase approximately linearly. Due to the linear increase, we see that as $tf_-/R_+$ goes to infinity, the entropy will then diverge. This is again expected since as the asymptotic time goes to infinity, the number of particles that are produced diverges (see Ref.~\cite{Green_rad}). This is a consequence of the fact that we keep the background fixed (i.e. ~$R_+$ is a constant). In reality, $R_+$ should decrease over time since the radiation is taking away mass and energy from the system. Therefore as $tf_-/R_+$ goes to infinity, the entropy of the entire system as measured by the asymptotic observer diverges as $R\rightarrow R_+$. 

\begin{figure}[bp]
\includegraphics{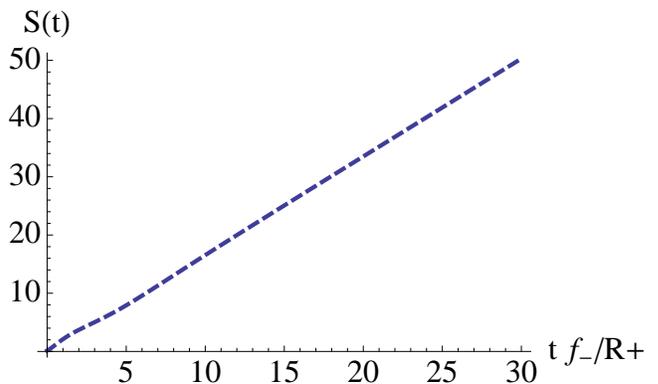}
\caption{We plot the entropy of the induced radiation created during the collapse as a function of $tf_-/R_+$. Except for the initial increase, the entropy increases approximately linearly as $tf_-/R_+$ increases.}
\label{EntR}
\end{figure}

Now we consider just the radiation which is induced during the collapse. In Figure \ref{EntR} we then plot the entropy as a function of $t_-/R_+$. Figure \ref{EntR} shows initially the entropy of the system is zero. Again, this is expected since initially the domain wall is in vacuum, meaning that initially there is no radiation being induced. Therefore the only degree of freedom is that of the domain wall, which then gives that the initial entropy of the radiation must be zero. As $tf_-/R_+$ increases, initially there is a rapid increase in the entropy, but again, the entropy then increases linearly as $tf_-/R_+$ increases. As in the case of the entire system, as $tf_-/R_+$ goes to infinity, the entropy of the radiation also diverges. This is expected since the induced radiation diverges as $R\rightarrow R_+$; hence as the domain wall approaches the horizon, the occupation number for the induced radiation diverges. 

\begin{figure}[htbp]
\includegraphics{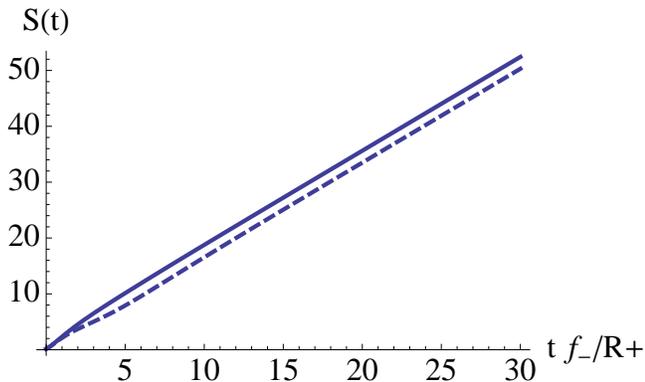}
\caption{We plot the entropy as a function of $tf_-/R_+$ for both the entire system and the induced radiation during the time of collapse. At the time $tf_-/R_+\approx7.5$, the slope of the two lines is approximately equal.}
\label{EntB}
\end{figure}

In Figure \ref{EntB} we plot the entropy as a function of asymptotic observer time $t$ of both the entire system and the induced radiation during the time of collapse. Figure \ref{EntB} shows that except the initial increase in the entropy, for later $tf_-/R_+$ (approximately $tf_-/R_+=7.5$), the slopes of the entropy versus time are approximately equal. Therefore, one can expect that the entropy of the domain wall is approximately constant for late times. 

\begin{figure}[htbp]
\includegraphics{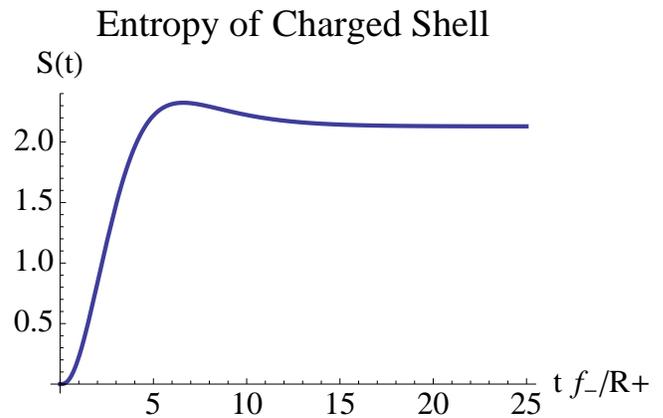}
\caption{We plot the entropy of the charged domain wall as a function of $tf_-/R_+$. Here we see for $tf_-/R_+>8$ the entropy is constant in time with a value of $S\approx 2(R_+/f_-)^2$.}
\label{Entdomain wall}
\end{figure}

What is of interest is the entropy of the collapsing charged domain wall, since this will form a black hole. To find the entropy of the charged domain wall, we can take the entropy of the entire system and subtract off the entropy of the induced radiation (since these are the only relevant objects which contribute to the entropy). The result is then given in Figure \ref{Entdomain wall}. We see some interesting features of the time-dependent entropy in Figure \ref{Entdomain wall}. Here we discuss these features; first we will make some general comments on the overall behavior of the time-dependent entropy, followed by discussions of late-time (values of $tf_-/R_+>8$) and the early-time (values of $t_-/R_+<8$) behaviors of the time-dependent entropy, respectively. 

Figure \ref{Entdomain wall} shows that initially the entropy of the charged domain wall is zero. As stated above, this is expected since initially there is only one degree of freedom. As $tf_-/R_+$ increases, the entropy of the domain wall rapidly increases. However, for late times ($tf_-/R_+>8$), the entropy of the charged domain wall goes to a constant. From the discussion above, this is expected since the late-time entropies for the entire system and for the radiation are approximately parallel. This feature is somewhat artificial though: the entropy is constant since we are assuming that the mass is approximately the Hamiltonian of the system, which is a constant of motion (see Ref.~\cite{WangGreenStoj}). This means that since we are holding the mass of the charged domain wall constant, we need to keep adding energy to the system to counteract the loss of mass (energy) from the radiation. Therefore one can expect that the entropy of the charged domain wall must be a constant for late times.

In reality, radiation takes mass away from the system, so the entropy of the charged domain wall will go to zero as $R_+$ goes to zero. Hence if the charged domain wall doesn't start off with enough mass, it could evaporate before ever creating the black hole. This means that after the charged domain wall disappears, all the entropy will go into the entropy of the radiation, since all the mass has dissipated.

For large values of $tf_-/R_+$ ($tf_-/R_+>8$) the entropy of the charged domain wall is constant. From Figure \ref{Entdomain wall} we see that our numerical value for the entropy of the charged domain wall for this region is
\begin{equation*}
  S\approx 2(R_+/f_-)^2.
\end{equation*}
Comparing with Eq.~(\ref{SBH}), we can view this discrepancy as a shift in the horizon radius $R_+$. In order to get the theoretical value for the entropy, Eq.~(\ref{SBH}), we see that we would require $R_+\rightarrow1.25R_s$. This is an understandable numerical error, which implies that our numerical solution is of the same order as the Hawking-Bekenstein entropy. This is in better agreement than the result for the Schwarzschild case, see Ref.~\cite{Green_ent}.

For small values of $tf_-/R_+$ ($tf_-/R_+<8$), the entropy of the charged domain wall seems to stop increasing around $tf_-/R_+\approx7.5$. This means that the entire change in entropy of the domain wall occurs in the region $0\leq tf_-/R_+\approx7.5$. At first this value for $tf_-/R_+$ seems arbitrary, however there is some physical significance to this time. 

First, we notice that the volume of the charged domain wall becomes approximately constant at $tf_-/R_+\approx7.5$. To illustrate this we will first require $R_0=nR_+$, where $n\in\Re$ ($\Re$ being the real numbers). From Eq.~(\ref{classical}) we can then write the position of the domain wall as
\bd
  R(t)=R_+\left(1+(n-1)e^{-tf_-/R_+}\right).
\ed
For illustration purposes we choose the value $n=10$, which at time $tf_-/R_+=7.5$ corresponds to a distance of $R\approx1.005R_+$ with velocity $\dot{R}=-f\approx-0.005$ (the numerical values for the position and velocity will be even less for smaller values of $n$). Hence the value of $tf_-/R_+\approx7.5$ corresponds to the region where the charged domain wall is very close to the horizon and the velocity of the charged domain wall, with respect to the asymptotic, goes approximately to zero. Therefore, as far as the asymptotic observer is concerned, the dynamical process of the charged domain wall collapsing is over and the domain wall has come to rest, meaning that there is now approximately a constant volume. From this point on, classically it takes an infinite amount of time for the domain wall to reach the horizon radius, hence travel the distance $R\approx0.005R_+$, see Section \ref{Model}. 

\begin{figure}[htbp]
\includegraphics{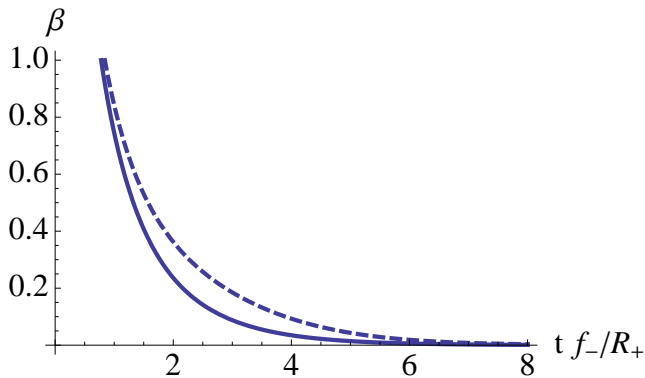}
\caption{We plot $\beta$ of the entire system and the induced radiation as a function of $tf_-/R_+$. Here we see for $tf_-/R_+\approx7.5$ the two values of $\beta$ become approximately equal, meaning the two systems are in thermal equilibrium.}
\label{beta}
\end{figure}

Second, we can show that the entire system and the induced radiation come into thermal equilibrium at this time. In Figure \ref{beta} we plot $\beta$ versus $tf_-/R_+$ for the entire system (continuous curve) and the induced radiation (dashed curve). Figure \ref{beta} shows that for the time $tf_-/R_+\approx7.5$ the values of the two $\beta$'s become approximately equal, meaning that the entire system and the induced radiation are now at the same temperature. Therefore the system is now in thermal equilibrium, meaning that there is no more change in entropy of the charged domain wall as $tf_-/R_+$ increases. Further more, the fluctuations (departure from thermality) in $\beta$ become smaller at this time, as discussed in Refs.~\cite{Green_rad}.

\begin{figure}[htbp]
\includegraphics{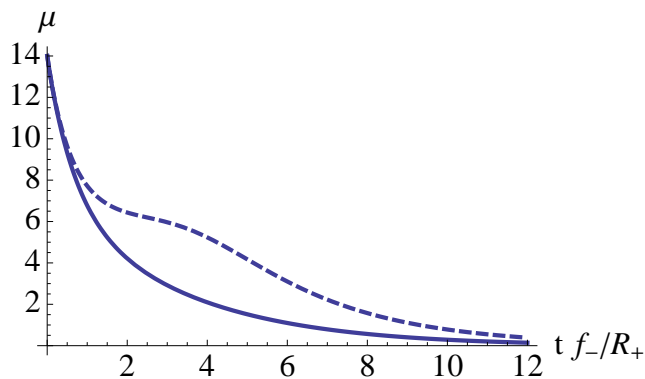}
\caption{We plot $\mu$ versus $tf_-/R_+$. The solid line corresponds to the entire system while the dashed line corresponds to the induced radiation only. Here we see that as $tf_-/R_+$ increases, the chemical potential for each goes to zero.}
\label{ChemPot}
\end{figure}

Finally we can evaluate the the chemical potential for both the entire system and for the induced radiation. From definition we can write the chemical potential as
\be
  \mu=\frac{\partial S}{\partial N}.
\ee
In Figure \ref{ChemPot} we plot the chemical potential for both the entire system and for the induced radiation. We can see that as $tf_-/R_+$ increases the chemical potential of the entire system and the induced radiation goes to zero. This means that the dispersion of particles goes to zero and the system goes into equilibrium. 

During the dynamical process, the entropy increases almost linearly, except for the bump or oscillation occurring during the time span $5<tf_-/R_+<10$. This oscillation may be attributed to several different circumstances. First, the oscillation may be caused by the non-thermal property of the radiation (see Ref.~\cite{Green_rad}). Secondly, this oscillation may be a manifestation of the error associated with the numerical calculations. Lastly, the oscillation may be an artifact of expanding the calculations beyond the region of validity, since we are using the near horizon approximation. Hence for values large compared to $R_+$, we cannot completely trust our result.

\section{Conclusion}
\label{Conclusion}

We studied the time evolution of the entropy of a collapsing spherically symmetric charged domain wall (representing a shell of charged matter) using the Functional Schr\"odinger and Louisville-von Neumann formalisms from the point of view of an asymptotic observer. This was done by investigating the change in entropy for the entire system (the charged domain wall and the induced radiation) and that of only the radiation induced during the collapse semi-classically, i.e. the background of the collapsing charged domain wall was treated classically while the radiation was treated quantum mechanically. As stated in Section \ref{Model}, this was done for the case $(GM)^2>Q^2$ only. The reason for this is that in the case $Q^2>(GM)^2$ Eq.~(\ref{classical}) contains an imaginary term in the exponential term. Therefore the solution contains sinusoidal terms, where the charged domain wall will oscillate at some radius $\tilde{R}$ outside the horizon radius. Thus, the horizon is never formed and a black hole never gets created.

To summarize, we have arrived at the following results. From Figure \ref{EntEntire} we see that $dS/dt>0$ for the entire system, which is in agreement with the second law of Thermodynamics. Second, from Figure \ref{EntR}, we see that initially the entropy of the radiation increases at a lower rate than that of the entropy of the entire system; however, for larger values of $tf_-/R_+$ the increases of each are approximately the same (as seen in Figure \ref{EntB}). The difference in the initial rates is due to the fact that each system evolves with a different time parameter ($\eta$ and $\tilde{\eta}$). Finally, from Figure \ref{Entdomain wall} we demonstrated that the entropy of the charged domain wall is constant for late times. As discussed in Section \ref{Entropy}, this is expected since we are holding the mass of the charged domain wall constant, thus we must keep adding energy to the system to maintain the mass of the charged domain wall. In a completely realistic model, this would not be the case. The entropy would be expected to decrease as $R_+$ decrease, since $R_+$ depends on the mass of the charged domain wall. So as the mass decreases the horizon radius would also decrease, meaning that the entropy of the charged domain wall would go to zero as $R_+$ goes to zero. Here we demonstrated that the late-time entropy (large values of $tf_-/R_+$) of the charged domain wall does in fact go to a constant as predicted by Bekenstein, see Refs.~\cite{Bekenstein,GibbonsHawking}. We found that the late time entropy of the charged domain wall is $S\approx2(R_+/f_-)^2$. By comparing Eq.~(\ref{SBH}) with our result, we see that our result is consistent with the accepted value for the entropy. The entropy for a collapsing massive shell was also derived in Ref.~\cite{Casadio}, where the authors used a mini-superspace effective action for the area and mass aspect of the shell. The authors show analytically that the entropy of the shell is on the order of the Bekenstein-Hawking entropy. 

Figure \ref{Entdomain wall} displays some other interesting features. First we see that all the change in entropy occurs during small values of $tf_-/R_+$, with the increase in entropy ending at a time $tf_-/R_+\approx7.5$. As discussed in Section \ref{Entropy}, this time is not an arbitrary time. To see this, first, for different values $R_0$, this time corresponds to when the charged domain wall is very close to the horizon and the corresponding velocity is very close to zero. From this point on the charged domain wall takes an infinite amount of time to traverse the remaining distance to the horizon. Thus, as far as the asymptotic observer is concerned, the dynamics of the charged domain wall has finished by the time $tf_-/R_+\approx7.5$, hence the observer won't see any further change in the entropy since the volume of the charged shell is approximately constant. Second, the ``inverse temperature" ($\beta$) for both the entire system and the induced radiation become equal at this time. This means that the two systems come into thermal equilibrium, as seen in Figure \ref{beta}. Furthermore, the fluctuations in the temperature become smaller by this time, as discussed in Ref.~\cite{Green_rad}. However, these oscillations may still be substantial and be the cause of the oscillation present from $5<tf_-/R_+<10$ (see Figure \ref{Entdomain wall}). Finally from Figure \ref{ChemPot} we see that the chemical potential for both the entire system and the induced radiation go to zero near this time. Again, this signifies that the dispersion of the induced particles goes to zero and the system arrives at equilibrium. 

\section*{Acknowledgements}

The author thanks Dejan Stojkovic for useful conversations and discussions.

\end{document}